# Novel room temperature Multiferroics for Random Access Memory Elements

**Ashok Kumar[1], Ram S. Katiyar[1], and James F. Scott[1,2]**

[1]Department of Physics and Institute for Functional Nanomaterials, University of Puerto Rico, San Juan, PR 00931-3343 USA

[2]Cavendish Laboratory, Dept. Physics, Cambridge University, Cambridge CB3 0HE, U. K.

*Abstract*—We have fabricated a variety of "PZT-PFW" $(PbZr_{0.52}Ti_{0.48}O_3)_{1-x}(PbFe_{2/3}W_{1/3}O_3)_x$ [PZTFWx; $0.2 < x < 0.4$] single-phase tetragonal ferroelectrics via chemical solution deposition (CSD) [polycrystalline] and pulsed laser deposition (PLD) [epitaxial] onto $Pt/Ti/SiO_2/Si(100)$ and $SrTiO_3/Si$ substrates. These exhibit ferroelectricity and (weak) ferromagnetism above room temperature with strain coupling via electrostriction and magnetostriction. Application of modest magnetic field strength ($\mu_0H < 1.0$ Tesla) destabilizes the long-range ferroelectric ordering and switches the polarization from ca. 22 $\mu C/cm^2$ (0.22 $C/m^2$) to zero (relaxor state). This offers the possibility of three-state logic (+P, 0, -P) and magnetically switched polarizations. Because the switching is of large magnitude (unlike the very small $nC/cm^2$ values in terbium manganites) and at room-temperature, commercial devices should be possible.

*Index Terms*—Magnetoelectrics; multiferroics; ferroelectrics; random access memories (RAMs).

∗Corresponding author

Email: rkatiyar@uprrp.edu (Prof. Ram S. Katiyar) and Email: jfs32@hermes.cam.ac.uk (Prof. James F Scott)

# I. INTRODUCTION

The development of practical magnetoelectric multiferroic devices for industry has concentrated on two embodiments: The first is for weak-field sensors for applications such as mine detection in harbors or related military devices. In this case practical devices favor sandwich-type bilayer or multilayer composites of ferroeloetrics such as lead zirconate-titanate (PZT) with magnets such as terphenyl-d. The second is for random access memory elements where the ability to switch magnetization with applied electric fields (and vice versa) would permit the combination of fast (possibly sub-nanosecond) WRITE operation with nondestructive magnetic READ operation. Such a combination would produce formidable competition for FLASH EEPROMs, particularly in view of the fact that magnetoelectric RAMs would operate at <1.0V, an international target for all microelectronics in the next decade which FLASH devices are unlikely to meet without cumbersome internal charge pumps [1]. However, most multiferroics to date (such as terbium manganites $TbMnO_3$ or $TbMn_2O_5$) switch only $nC/cm^2$ with applied magnetic fields – approximately 1000x too small for reliable discrimination between "1" and "0" states [2, 3]. Moreover, they operate only at cryogenic temperatures. $BiFeO_3$ might operate at room temperature, but its magnetism is very weak (ca. 0.03 emu/cc) and it neither permit switching large magnetizations with electric fields nor large polarizations with magnetic fields [4, 5, 6]. Clearly a completely new line of thinking is required to overcome the intrinsic limitations of direct linear magnetoelectric (ME) coupling of form PM in the free energy, where P is polarization and M, magnetizion. In very recent work we have established theoretically [7] and experimentally [8] that single-phase tetragonal crystals combining $PbFe_{2/3}W_{1/3}O_3$ (PFW) with $PbZr_{0.52}Ti_{0.48}O_3$ (PZT) have polarization P and magnetization M strongly coupled indirectly via strain, viz. electrostriction and magnetostriction. For PFW/PZT ratios of from 1:4 to 2:3 these materials are borderline relaxors, and the transition to long-range ferroelectricity occurs near room temperature. As a result, modest applied magnetic fields can destabilize the long-range electrical order and reduce P from large values (0.30-0.70 $C/m^2$) to zero. This comes about by changing the electrical polarization relaxation time from 200 ns to ca. 100 μs. [7, 9, 10].

In the present work we extend earlier studies of sol-gel chemical solution deposited (CSD) polycrystalline films to pulsed laser deposited (PLD) epitaxial films and also give detailed device performance for PFWFWx on $SrTiO_3/Si$ substrates. The new studies include band match-ups, capacitance versus frequency data at -8.0V gate voltage, memory windows (as large as 3.0V), and current-voltage J(V) curves.

## II. BAND MATCH-UPS

Fig.1 illustrates the energy level diagram (energy versus real space through the heterojunction device) for Pt top–electroded PZTFWx on $SrTiO_3$ (6 nm) with a p-Si(111) substrate. Here Φ is the Pt work function 5.3 eV; χ, the semiconductor electron affinity (3.5 eV for PZTFWx with 80% PZT; 4.15 eV for $SrTiO_3$; and 4.05 eV for Si); $E_g$ the bandgaps (3.4 eV for PZTFWx; 3.3 eV for $SrTiO_3$; and 1.1 eV for Si); $E_v$ and $E_c$, the valence and conduction band edges [11,12,13]. In the diagram the vacuum level has been allowed (due to charge injection) to differ slightly in each material; this change of ca. 0.3 eV is to match the known Schottky barrier height of 0.20 eV and 0.46 eV for well-annealed $600^oC$ and $650^oC$ $SrTiO_3$ on Si respectively [13, 14]. The 2.5 eV gap for Si is the Schottky barrier for hole injection from $SrTiO_3$ into Si. Depending on the surface treatments, valence band off-sets vary from 2.38 to 2.64 for $SrTiO_3/Si$ hetrostructures. We took the average of these values i.e. 2.5 eV in Fig.1. Fig 1 indicates that the conduction band minimum of $SrTiO_3$ is below that of semiconductor, a situation known as negative conduction band off-set. The work function of Pt is 5.3 eV The electron affinity and band gap of PZTFWx are 3.4 eV and 3.5 eV respectively and imply that the barrier height at Pt/PZTFWx interface is 1.8 eV for electrons and 1.6 eV for holes; but the hole injection into the $SrTiO_3$ is precluded by an additional 0.55 eV at the PZTFWx/$SrTiO_3$ interface and another 2.0 eV at the $SrTiO_3$/Si interface, so that conduction through the entire heterojunction combination should be n-type. This high band off-set at the Pt/PZTFWx interface limits the leakage currents for negative applied gate voltage, whereas the band off-set across the $SrTiO_3$/Si interface is in itself insufficient to protect the leakage currents. Despite these facts, our Metal Multiferroic Insulator Semiconductor (MMIS) structure showed great future for memory applications.

## III. EXPERIMENTAL DETAILS

To fabricate the MMIS structure, 6 nm $SrTiO_3$ was deposited on a p-type Si(111) substrate by PLD techniques. Initially we tried to deposit amorphous $SrTiO_3$ at low temperatures to realize the structure, but this yielded much higher leakage currents, which in turn gives unsaturated accumulation regions under the application of negative bias voltage. So instead crystalline $SrTiO_3$ (6 nm) was grown at $700^0C$ with 100 mT of oxygen atmosphere. Three different compositions of PZTFWx (0.20<x<0.40) films with varying deposition temperature were fabricated by PLD employing a KrF excimer laser ($\lambda$=240 nm). The PZTFWx layer was then deposited on the ST layer at $600^oC$ under an oxygen pressure of 200mTorr, using a laser energy density of (1.5 J/cm$^2$) and repetition rate of 10 Hz. After deposition the PZTFWx was annealed at $600^oC$ for 30 minutes in oxygen at a pressure of 300Torr. Finally the films were cooled down to room temperature at a slow rate. The total thicknesses of the films were around 350 nm. The structural analysis was done with a Siemens D500 x-ray diffractometer (Cu K$\alpha$ radiation) in a $\theta$- 2θ scan. Atomic force microscopy (Veeco-AFM-contact mode) was used to examine the morphology and the surface roughness. The film thickness was determined using an X-P-200 profilometer. DC sputtering was carried out for depositing the Pt top electrode of 3.1 x $10^{-4}$ $cm^2$ area using a shadow mask. The dielectric properties and capacitance-voltage C(V) in the frequency range of 100 Hz to 1 MHz were studied using an impedance analyzer HP4294A (Agilent Technology Inc.). Current-voltage I(V) measurement was carried out utilizing a Keithley electrometer model 6517A at 296 K. I(V) datas were recorded with a step time of 10 second and delay time 30 second from sweep electric field of -300 kV/cm to 300 kV/cm.

## IV.  XRD STRUCTURE AND AFM TOPOGRAPHY

Fig.2 (a) shows the X-ray intensities versus 2θ for the polycrystalline PZTFWx films of: (i) x= 0.20; (ii) x= 0.30; (iii) x= 0.40 (top to bottom) thin films on high k-SrTiO$_3$ (6-nm) coated p-Si (111). The data imply a good single-phase tetragonal material of perovskite structure. Details of the crystal structures are given elsewhere [10]. Surface morphology of all these compositions was thoroughly studied in order to find out their real practical applications. Figs.2(b),(c),(d) illustrate the surface topography of the films, which indicates well-defined grains with an average size 40-100 nm; most of the grain size was in the range of 40–50 nm with some agglomeration of larger grains. The average surface roughness of all the compositions lies between 3-6 nm. The average surface roughness of the films decreases with increase in Fe and W compositions. Fig. 2(d) reveals a darker reddish area with low z-height (< 10 nm) prominent in 40:60% FeW/TiZr compositions. It also suggests the presence of ordered regions -- i.e. polar nano-regions (PNRs) -- in a disordered matrix. These PNRs contribute to the enhancement in dielectric dispersion and decrease in polarization. Some of the fine grains agglomerate due to higher fabrication temperature, which in turn gives the bigger grains seen in the AFM picture. The observed larger grain size and higher surface roughness may be due to growth at higher temperature and utilization of the high energy PLD process. PZTFWx (x =0.20) films showed well distributed homogenously packed grains throughout the surface; for higher PFW compositions more inhomogenous surfaces were observed.

## V.  FREQUENCY DEPENDENCE OF CAPACITANCE

Fig.3 illustrates capacitance and dielectric loss tangent versus frequency from 100 Hz to 1 MHz at -8.0V gate voltage. The capacitance observed from 300 to 400 pF corresponds for the geometry used to a dielectric constant of ca. 220, and hence to a dielectric loss of between 20 and 65 (10-30% as frequency increases from 10 kHz to 1 MHz).  The loss could be a problem for high clock-rate RAM devices. The high loss and high dielectric dispersion was observed at higher frequencies can be explained as arising from two sources: (i) the presence of a resonance in the test circuit; (ii) PNRs and related phenomena -- i.e. high loss for high frequencies and low loss for low frequencies.  We think the latter are dominant.

## VI.  MEMORY WINDOW

Figs.4 (a), (b), (c) show memory window data as functions of sweep voltage at 1 MHz operating frequency.  Curve (a) is for 20% PFW; (b) is 30% PFW; (c) is 40% PFW. A large window of ca. 4.0V is observed for the 40% PFW sample The C(V) curve is similar to that of a normal metal-oxide-semiconductor (MOS) structure, clearly showing regions of accumulation, depletion, and inversion. It should be remember that all these compositions show very large polarization and low coercive field, as presented in our earlier reports. The figure displays large clockwise C(V) hysteresis and a memory window controlled by ferroelectric hysteresis of the PZTFWx on the Si surface. Figure 5 illustrates the memory window of PZTFWx as function of sweep voltage with an almost negligible memory window below -3 V sweep voltage, saturated C(V) hysteresis until -10 V, above which gate voltage the MMIS structure started leaking during the DOWN sweep, whereas during the UP sweep (in gate voltage) accumulation regions showed good saturation. These observations indicate that the materials have high potential for RAM applications. The band match-up illustrates that the existing MMIS structures have a very high band off-set value (1.8 eV for electrons and 1.6 eV for holes) with a negative conduction band off-set (0.3V) at the SrTiO$_3$/Si interface. The experimentally observed leakage behavior of the DOWN sweep is well supported by a simple energy-band diagram. The observed memory window is almost frequency independent. A saturated memory window in the present structures cannot be obtained due to injection of charge carriers (electrons) during the DOWN voltage sweep.

## VII. CONDUCTIVITY

In Fig.6 we exhibit current density-applied electrical field J(E) data. The specimens, as in earlier figures, are: (a) 20% PFW; (b) 30% PFW; (c) 40% PFW. All these specimens showed very low leakage current for this MMIS structure; these leakage currents are higher if we applied the electric field from the semiconductor p-Si side. The different specimens showed the same characteristic J(E) behavior of with very low leakage ~ $10^{-5}$ to $10^{-7}$ A/cm$^2$ ($10^{-9}$ to $10^{-11}$ A/m$^2$) under -10V gate voltage. Due to low leakage behavior, it is hard to distinguish leakage current differences among them; but x= 0.30 % of PFW has less leakage. Fig. 7(a) shows ln (J/E) versus E$^{1/2}$ graphs (-5V negative bias accumulation) for all the compositions. For compositions x < 0.30, it illustrates constant behavior with very low slope value ($\beta/k_BT$ = 0.12-0.21) and linearity (for x

= 0.40; β/k$_B$T = 0.85) at higher compositions suggesting the conduction mechanism either due to Poole–Frenkel (PF) or Schottky emission (SE) [15,16]. The equations mentioned below are used to calculate the conduction mechanism of MMIS structure.

$$J / E = J_0 \exp\left(\frac{\beta E^{1/2}}{\kappa_B T}\right)$$ ………………………………………………………………(1)

$$\text{where } \beta = \left(\frac{e^3}{\alpha \pi \varepsilon_0 \varepsilon_\infty}\right)^{1/2}$$ ………………………………………………………(2)

Here $J_0$ is a constant; E, electric field; $k_B$, Boltzmann's constant; T, absolute temperature; β is a constant depending upon conduction mechanism assumed; $\varepsilon_\infty$, the high frequency dielectric constant; and α = 1 for PF and 4 for SE conduction mechanisms. We have fitted our experimental data for x= 0.20 (as shown in Fig. 7 (b)) and x= 0.40 (as shown in Fig. 7 (c)); the calculated values are β~ 0.97 x 10$^{-21}$ and 3.43 x 10$^{-21}$ respectively.  Theoretically the Poole–Frenkel (PF) model gives of β~ 0.88 x 10$^{-21}$, which matches the experimental value within 10% for the x = 0.20 specimen, showing that leakage current for x < 0.30 compositions is bulk-limited. For x> 0.40 the experimental value of β is somewhat larger, suggesting that  the leakage current is due to both electrode- limited (SE) and bulk-limited (PF) conduction mechanisms.

<div style="text-align:center">

CONCLUSION

</div>

MMIS device structures of Pt/PZTFWx/SrTiO$_3$/Si (111) are fabricated for different compositions of PFW: (a) x= 0.20, (b) x= 0.30, (c) x= 0.40. All these composition showed well behaved and saturated C(V) hysteresis with 3-4 V memory windows, suggesting good candidates for multiferroic RAMs. J(E) characteristics of these specimens illustrate low leakage currents at room temperature dominated by bulk-limited n-type Poole-Frenkel conduction. Energy band match-up diagram and experimental results suggest that all these specimens might provide a wide memory window and good DOWN saturation if used with alternative high band-gap and high k-dielectric materials instead of $SrTiO_3$. Since all the compounds are multiferroic in nature, under the application of external magnetic field we might destabilize the saturated gate voltage and produce more than two logic states with additional functionalities.

### REFERENCES:

[1] James F. Scott and Carlos A. Paz de Araujo, **"Ferroelectric Memories,"** *Science*, Vol. 246. no. 4936, pp. 1400 – 1405, December 1989.

[2] T. Kimura, T. Goto, H. Shintani, K. Ishizaka, T. Arima & Y. Tokura, **"Magnetic control of ferroelectric polarization,"** *Nature,* **426**, 55-58, (2003).

[3] M. Fiebig, Th. Lottermoser, D. Fröhlich, A. V. Goltsev & R. V. Pisarev, "Observation of coupled magnetic and electric domain," *Nature,* **419**, 818-820, (2002).

[4] W. Eerenstein, N. D. Mathur, & J. F. Scott, "Multiferroic and magnetoelectric materials," *Nature,* **442**, 759-764, (2006).

[5] J. Wang, J. B. Neaton, H. Zheng, V. Nagarajan, S. B. Ogale, B. Liu, D. Viehland, V. Vaithyanathan, D. G. Schlom, U. V. Waghmare, N. A. Spaldin, K. M. Rabe, M. Wuttig, R. Ramesh "Epitaxial $BiFeO_3$ Multiferroic Thin Film Heterostructures," *Science,* **299**, 1719, (2003).

[6] W. Eerenstein, F. D. Morrison, J. Dho, M. G. Blamire, J. F. Scott, and N. D. Mathur, "Comment on Epitaxial $BiFeO_3$ multiferroic thin film heterostructures," *Science*, **419**, 1203a, (2005).

[7] Ashok Kumar, G. L. Sharma, R.S. Katiyar, J. F. Scott, R. Pirc, R. Blinc "Magnetic control of large room-temperature polarization," *J. Phys. Condens. Matter* **21**, 382204, (2009).

[8] R. Pirc, R. Blinc, and J. F. Scott, "Mesoscopic model of a system possessing both relaxor ferroelectric and relaxor ferromagnetic properties," *Physical Review* B **79**, 214114, (2009).

[9] Ashok Kumar, R.S. Katiyar, J. F. Scott, "Positive temperature coefficient of resistivity and negative differential resistivity in lead iron tunstate-lead zirconate titate," *Applied Physics Letters* **94**, 212903, (2009).

[10]   Ashok Kumar, R.S. Katiyar, J. F. Scott, Journal of *Applied Physics (submitted).*

[11] J. Senzaki, K. Kurihara, N. Nomura, O. Mitsunaga, Y. Iwasaki, and T. Ueno, "Characterization of Pb(Zr, Ti)O$_3$ Thin Films on Si Substrates Using MgO Intermediate Layer for Metal/Ferroelectric/Insulator/Semiconductor Field Effect Transistor Devices**,"** *Jpn. J. Appl. Phys.*, Part1, **37**, 5150, (1998).

[12]   N. M. Murari, R. Thomas, S. P. Pavunny, J. R. Calzada, and R. S. Katiyar, "DyScO$_3$ buffer layer for a performing metal-ferroelectric-insulator-semiconductor structure with multiferroic BiFeO$_3$ thin film," *Appl. Phys. Lett.*, **94**, 142907, (2009).

[13]   F. Amy, A. S. Wan, A. Kahn F. J. Walker and R. A. McKee, "Band offsets at heterojunctions between SrTiO$_3$ and BaTiO$_3$ and Si(100)," J. *Appl. Phys.*, **96**, 1635, (2004).

[14]   R. A. McKee, F. J. Walker, M. Buongiorno Nardelli, W. A. Shelton, G. M. Stocks, "The Interface Phase and the Schottky Barrier for Crystalline Dielectric on Silicon," *Science*, **300**, 1726 (2003).

[15]    J. G. Simmons, "Poole-Frenkel Effect and Schottky Effect in Metal-Insulator-Metal Systems," *Phys. Rev.* **155**, 657, (1967).

[16]   N. Izyumskaya , Y. -I. Alivov, S. -J. Cho, H. Morkoç, H.  Lee, Y. -S. Kang , "Processing, Structure, Properties, and Applications of PZT Thin Films" *Critical Reviews in Solid State and Materials Sciences*, **32**, 111–202, 2007.

**Acknowledgements:**

Manuscript received Jan 6, 2010. This work was partially supported by DOD W911NF-05-1-0340, W911NF-06-1-0030, W911NF-06-1-018, DoE FG 02-08ER46526 grants to UPR, and EU STREP Multiceral funding at Cambridge.

Ashok Kumar & Ram S Katiyar : Author is with the Department of Physics and Institute for Functional Nanomaterials, University of Puerto Rico, San Juan, PR 00931-3343 USA (Fax: 787-764-2571) (e-mail: ashok553@gmail.com & ashok.kumar@uprrp.edu).

James F Scott is with Cavendish Laboratory, Dept. Physics, Cambridge University, Cambridge CB3 0HE, U. K. (e-mail: jfs32@hermes.cam.ac.uk) and also with the Department of Physics and Institute for Functional Nanomaterials, University of Puerto Rico, San Juan, PR 00931-3343 USA.

Ram S Katiyar : Author is with the Department of Physics and Institute for Functional Nanomaterials, University of Puerto Rico, San Juan, PR 00931-3343 USA (Fax: 787-764-2571). (e-mail: rkatiyar@uprrp.edu)

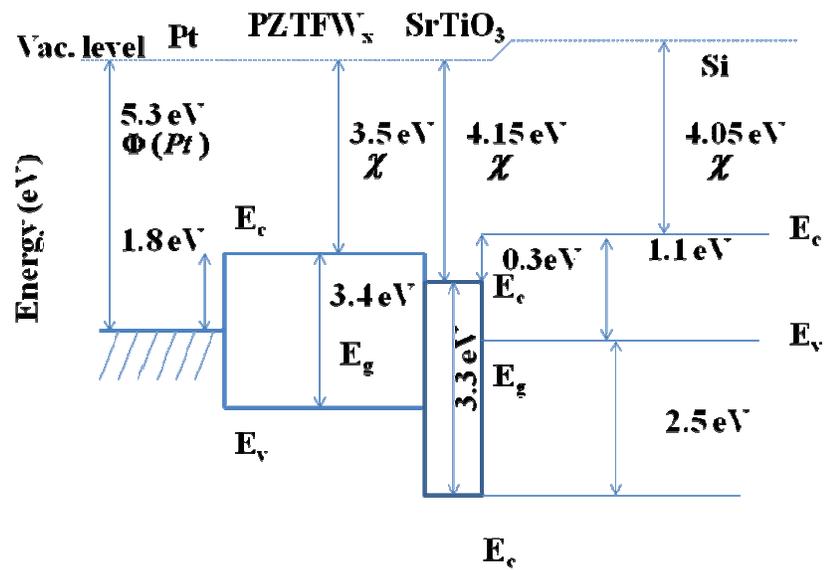

**Fig. 1 Summary of band alignment and interface of Pt/PZTFW$_x$/SrTiO$_3$/Si**

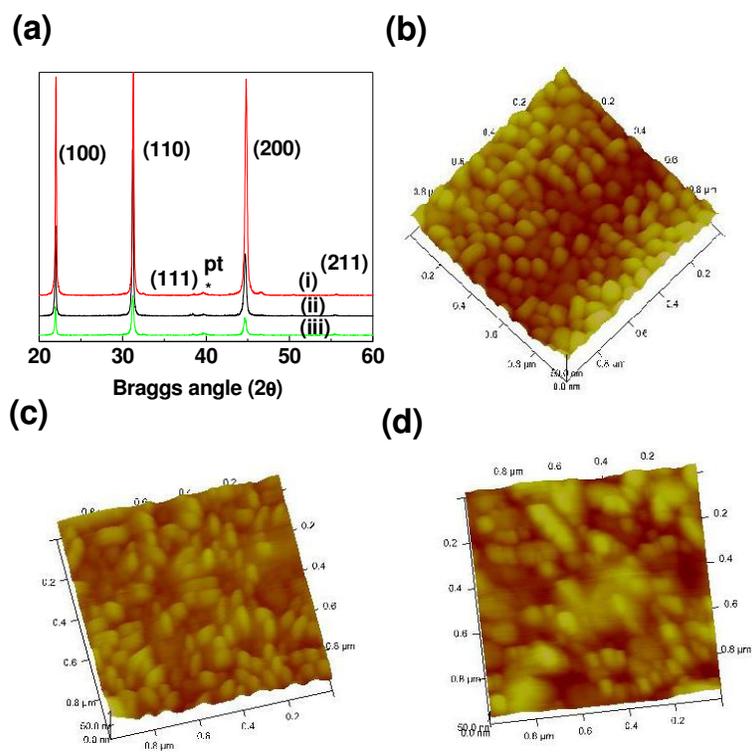

Fig. 2 (a) Room-temperature XRD of the PZTFWx: thin films:
Top to bottom (a) x= 0.20 (i), 0.30 (ii), 0.40 (iii); (b) surface
topography for x= 0.20 (c) for x=0.30 (d) for x=0.40 across 1µm
x 1µm x 50 nm (z-scale).

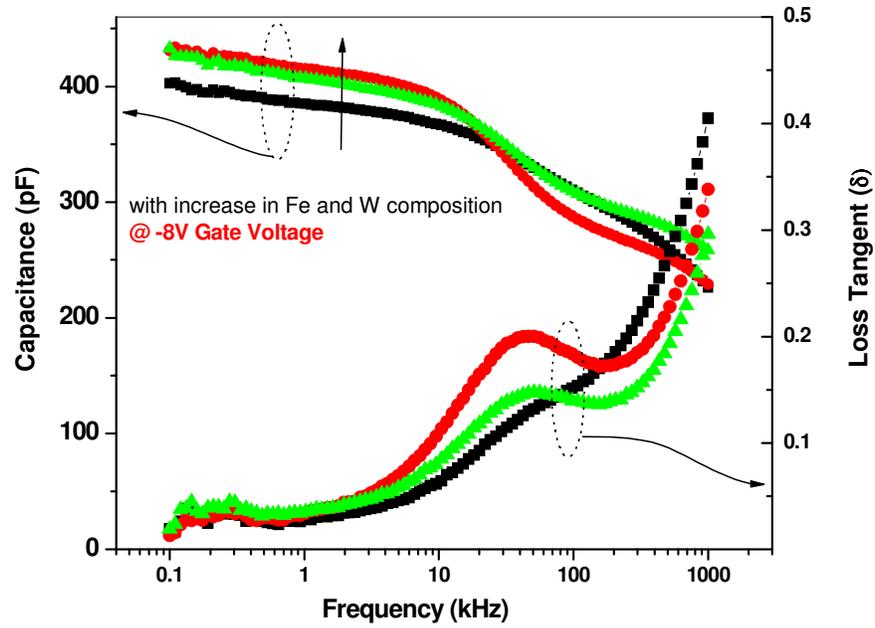

Fig. 3 Room temperature capacitance and tangent loss behavior as function of experimental frequency window from 100Hz to 1MHz for all the compositions at -8V applied gate potential.

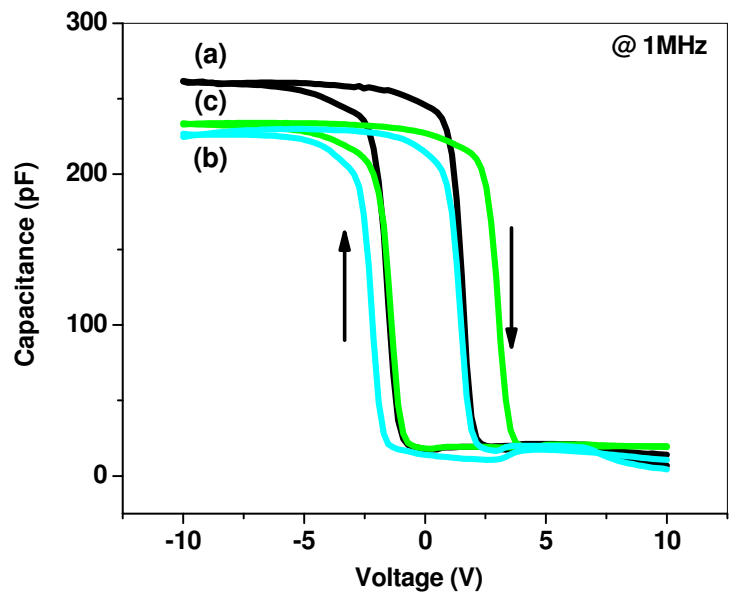

Fig. 4 High frequency C(V) characteristics of Pt/PZTFWx/SrTiO$_3$/(111)Si MMIF structures for different compositions of PFWx: (a) x= 0.20, (b) x= 0.30, (c) x= 0.40.

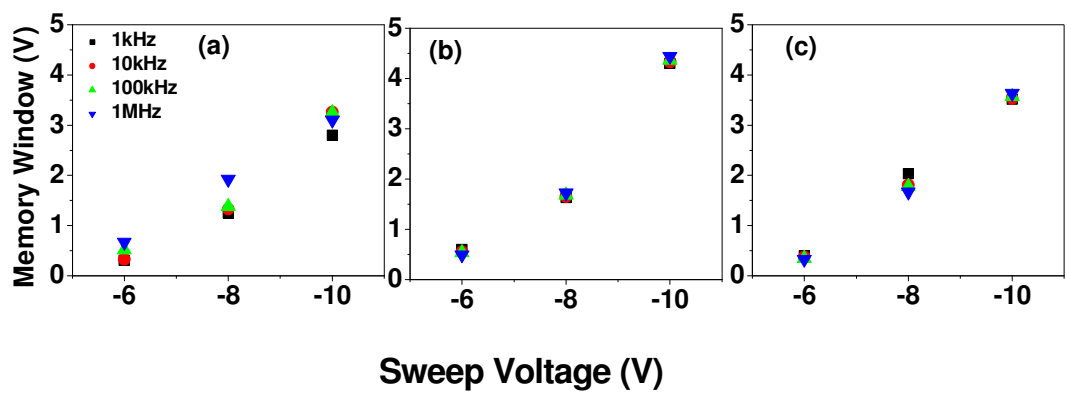

Fig. 5 Memory window of Pt/PZTFWx/SrTiO₃/Si (111) MMIF structure as function of sweep gate voltage for different compositions of PFW (a) x= 0.20, (b) x= 0.30, (c) x= 0.40 at various frequencies. Memory window values were as high as 3-4 V for all these compositions.

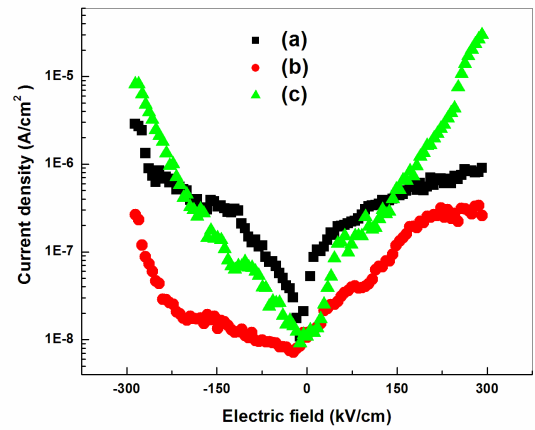

Fig. 6 Room temperature J(E) characteristic of Pt/PZTFWx/SrTiO$_3$/(111)Si MMIF structures as function of sweep gate field for different compositions of PFW (a) x= 0.20, (b) x= 0.30, (c) x= 0.40. All these compositions showed very low leakage current at room temperature with -300kV/cm (-30MV/m) gate field.

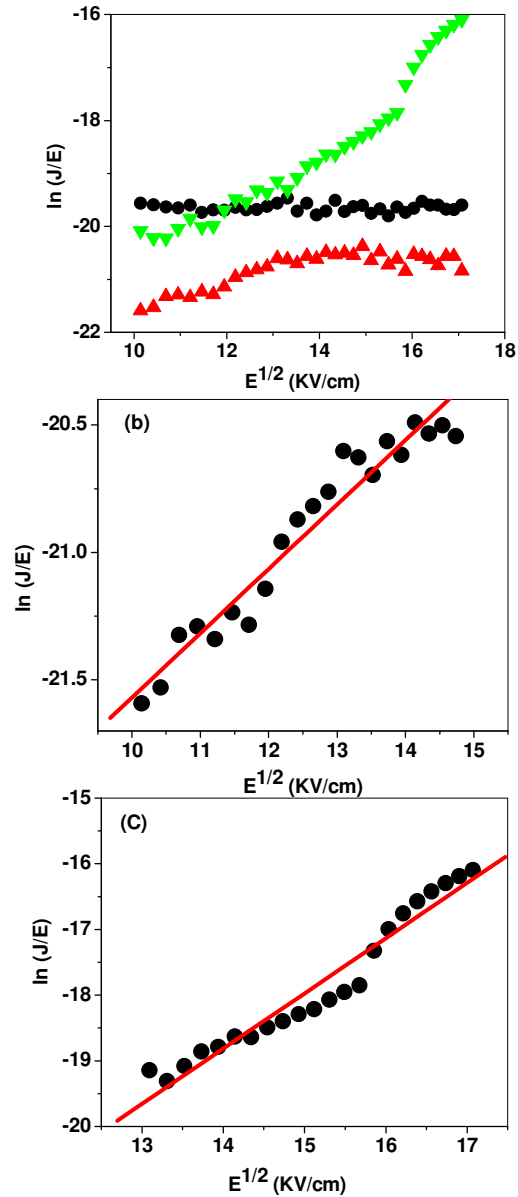

Fig. 7 Room temperature J(E) characteristics of Pt/PZTFWx/SrTiO$_3$/(111)Si MMIF structures as function of sweep gate field > 150 kV/cm (> 15MV/m) for different compositions of PFW: (a) for all the compositions; (b) x= 0.20 (at certain higher electric fields), (c) x= 0.40 (for higher fields). All these compositions showed almost negligible temperature dependent leakage currents for low gate voltage. Red line (solid line) indicates linear fitting of experimental data.